\RequirePackage{amsmath}
\documentclass{article}

\usepackage[left=3cm, right=3cm, top=2cm, bottom = 2cm]{geometry}

\usepackage{lmodern}
\usepackage[UKenglish]{isodate}

\usepackage{xcolor}
\usepackage{textcomp}

\usepackage{cite}

\usepackage[T1]{fontenc}
\usepackage{graphicx}

\usepackage[utf8]{inputenc}
\usepackage{stfloats}

\usepackage[hidelinks]{hyperref}
\usepackage{listings}
\usepackage{xspace}
\usepackage[ruled,vlined,linesnumbered]{algorithm2e}
\usepackage{doi}

\usepackage{array,booktabs}
\usepackage{tabu}
\usepackage{xparse} 
\usepackage{glossaries} 
\usepackage{subfiles} 
\usepackage{listings} 
\usepackage{newfloat} 
\usepackage{verbatim} 
\usepackage{fancyvrb} 
\usepackage[misc,geometry]{ifsym}

\usepackage{regexpatch}
\usepackage[textsize=normalsize,textwidth=3cm]{todonotes}

\makeatletter
\xpatchcmd{\@todo}{\setkeys{todonotes}{#1}}{\setkeys{todonotes}{inline,#1}}{}{}
\makeatother

\newcommand{\q}[1]{\vskip0.5em\textbf{#1}}

\DeclareFloatingEnvironment[fileext=frm,placement={!ht},name=Listing]{listing}

\SetKwProg{Fn}{Plan}{}{}

\usepackage{scalerel}
\usepackage{tikz}
\usetikzlibrary{svg.path}

\definecolor{orcidlogocol}{HTML}{A6CE39}
\tikzset{
  orcidlogo/.pic={
    \fill[orcidlogocol] svg{M256,128c0,70.7-57.3,128-128,128C57.3,256,0,198.7,0,128C0,57.3,57.3,0,128,0C198.7,0,256,57.3,256,128z};
    \fill[white] svg{M86.3,186.2H70.9V79.1h15.4v48.4V186.2z}
                 svg{M108.9,79.1h41.6c39.6,0,57,28.3,57,53.6c0,27.5-21.5,53.6-56.8,53.6h-41.8V79.1z M124.3,172.4h24.5c34.9,0,42.9-26.5,42.9-39.7c0-21.5-13.7-39.7-43.7-39.7h-23.7V172.4z}
                 svg{M88.7,56.8c0,5.5-4.5,10.1-10.1,10.1c-5.6,0-10.1-4.6-10.1-10.1c0-5.6,4.5-10.1,10.1-10.1C84.2,46.7,88.7,51.3,88.7,56.8z};
  }
}

\newcommand\orcidicon[1]{\href{https://orcid.org/#1}{\mbox{\scalerel*{
\begin{tikzpicture}[yscale=-1,transform shape]
\pic{orcidlogo};
\end{tikzpicture}
}{|}}}}

\newglossaryentry{mas:jacamo}{name=JaCaMo,description={a platform for the development of multi-agent systems}}
\newglossaryentry{mas:jason}{name=Jason,description={AgentSpeak interpreter for programming agents}}
\newglossaryentry{mas:cartago}{name=CArtAgO,description={}}
\newglossaryentry{mas:moise}{name=\ensuremath{\mathcal{M}\textsc{oise}},description={}}
\newacronym{cnp}{CNP}{Contract Net Protocol}
\newacronym{mas}{MAS}{Multi-Agent System}
\newacronym{bdi}{BDI}{Belief Desire Intention}

\newacronym{lfc}{LFC}{Liverpool Formidable Constructors}
\newacronym{mlfc}{MLFC}{Manchester and Liverpool Formidable Constructors}
\newglossaryentry{fit}{name=FIT BUT,description={}}
\newglossaryentry{trg}{name=TRG,description={}}
\newglossaryentry{dtu}{name=GOAL-DTU,description={}}
\newglossaryentry{jac}{name=JaCaMo Builders,description={}}
\newglossaryentry{lti}{name=LTI-USP,description={}}

\newacronym{mapc}{MAPC}{Multi-Agent Programming Contest}

\newglossaryentry{team:teamart}{name=\texttt{TeamArtifact},description={}}
\newglossaryentry{team:agentart}{name=\texttt{EISArtifact},description={}}
\newglossaryentry{team:translator}{name=\texttt{Translator},description={}}

\newacronym{fd}{FD}{Fast Downward}

\newacronym{csp}{CSP}{Communicating Sequential Processes}
\newacronym{fdr}{FDR}{Failures-Divergences Refinement}

\input{./snippet-code/jacamo.pygstyle}

\begin{document}

\title{MLFC: From 10 to 50 Planners in the Multi-Agent Programming Contest\thanks{Work supported by UK Research and Innovation, and EPSRC Hubs for ``Robotics and AI in Hazardous Environments'': EP/R026092 (FAIR-SPACE) and EP/R026084 (RAIN). Cardoso's work is also supported by Royal Academy of Engineering under the Chairs in Emerging Technologies scheme.}}


\author{Rafael C. Cardoso \orcidicon{0000-0001-6666-6954} \\ 
\small{\parbox{7cm}{\centering The University of Manchester, \\Department of Computer Science,\\ Manchester, UK}}  
\and
Angelo Ferrando \orcidicon{0000-0002-8711-4670} \\ 
\small{\parbox{7cm}{\centering University of Genova,\\ Department of Computer Science, Bioengineering, Robotics and Systems Engineering (DIBRIS), Genova, Italy}}
\and
Fabio Papacchini \orcidicon{0000-0002-0310-7378} \\ 
\small{\parbox{7cm}{\centering University of Liverpool,\\ Department of Computer Science,\\ Liverpool, UK}}
\and 
Matt Luckcuck \orcidicon{0000-0002-6444-9312} \\ 
\small{\parbox{7cm}{\centering Maynooth University,\\ Department of Computer Science,\\ Maynooth, Ireland}}
\and 
Sven Linker \orcidicon{0000-0003-2913-7943} \\ 
\small{\parbox{7cm}{\centering Lancaster University in Leipzig,\\ School of Computing and Communications,\\ Leipzig, Germany}}
\and 
Terry R. Payne \orcidicon{0000-0002-0106-8731} \\ 
\small{\parbox{7cm}{\centering University of Liverpool,\\ Department of Computer Science,\\ Liverpool, UK}}
}

%

\maketitle

\begin{abstract}
\noindent In this paper, we describe the strategies used by our team, MLFC, that led us to achieve the 2$^{nd}$ place in the 15$^{th}$ edition of the Multi-Agent Programming Contest. The scenario used in the contest is an extension of the previous edition (14$^{th}$) ``Agents Assemble'' wherein two teams of agents move around a 2D grid and compete to assemble complex block structures. We discuss the languages and tools used during the development of our team. Then, we summarise the main strategies that were carried over from our previous participation in the 14$^{th}$ edition and list the limitations (if any) of using these strategies in the latest contest edition. We also developed new strategies that were made specifically for the extended scenario: cartography (determining the size of the map); formal verification of the map merging protocol (to provide assurances that it works when increasing the number of agents); plan cache (efficiently scaling the number of planners); task achievement (forming groups of agents to achieve tasks); and bullies (agents that focus on stopping agents from the opposing team). Finally, we give a brief overview of our performance in the contest and discuss what we believe were our shortcomings.

\end{abstract}

\section{Introduction} 
\label{sec:intro}

\begin{sloppypar}
In this paper, we focus on the strategies used by our team, the \acrfull{mlfc}. We provide some context and basic concepts of the scenario and the simulation environment; however, we assume the reader has some prior knowledge about the \acrfull{mapc}\footnote{\url{https://multiagentcontest.org/}} in order to better understand our contributions.
\end{sloppypar}

The 15$^{th}$ edition of the MAPC\footnote{\url{https://multiagentcontest.org/2020/}} is part of an annual competition in the area of multi-agent programming. This latest edition extended the ``Agents Assemble'' scenario from the 14$^{th}$ MAPC~\cite{MAPC2019}. The main goal of the scenario remained the same: two teams of agents compete in a 2D grid to fulfil tasks that are randomly generated by the simulation server, and those teams that complete tasks successfully receive some currency, with the aim of maximising this currency.
These tasks comprise assembling and delivering complex block structures of random size, and possibly requiring blocks of different types. A match starts once both teams have connected to the server, and then it proceeds synchronously with each simulation step having a 4 second timeout for agents to submit their actions for that step. A match has 3 rounds, with each round having a random configuration of the servers parameters (different grids, task size, etc.).

The most notable changes to the latest edition of the MAPC include:
\begin{enumerate}
    \item A circular (i.e., borderless and continuous) grid map; e.g., in a 50x50 grid, if an agent at cell 0,0 (i.e., in the top left corner) moves up, it would arrive in cell 0,49 (i.e., the bottom left corner);
    \item An increase in the number of agents for each round: 15, 30, and 50 for rounds 1, 2, and 3 respectively (previously this was 10 for all rounds);
    \item The addition of a new type of facility called \emph{taskboards}, such that an agent is required to be near a taskboard to accept a task.  The same agent must also submit the completed task to the server (previously tasks did not need to be accepted and any agent could submit the task).
\end{enumerate}

Our team, MLFC\footnote{The source code for MLFC is available and can be downloaded from the following URL: \url{https://github.com/autonomy-and-verification-uol/mapc2020-lfc}}, achieved second place in the 15$^{th}$ \acrshort{mapc}. 
A prior incarnation of the team, \acrfull{lfc}, had participated in the previous edition of the competition and won first place. 
We therefore recapitulate the main strategies used in that edition that were also useful in the new edition of the contest, although for a more comprehensive explanation we refer the reader to~\cite{Cardoso20d}, which has a complete description of the approach taken by LFC. The main contribution of this work is in the description of the new strategies that were developed specifically for the 15$^{th}$ \acrshort{mapc}. These strategies include: 
\emph{cartography} -- this tackles extension (1) to make our agents capable of discovering the size of the map (information that is unknown during a match); 
\emph{verification of map merging} -- in order to increase the confidence that our previous strategy for merging map information works well when scaling the number of agents, we performed a formal verification of our protocol (thus addressing extension (2)); 
\emph{plan cache} -- our previous strategy of using automated planners to plan the movement of our agents in the grid had to be adapted to cope with extension (2); 
\emph{achieving tasks} -- small changes were necessary to address extension (3), but a number of more significant modifications were required to make efficiency improvements necessary due to extension (2); and
\emph{bullies} -- the additional agents from extension (2) allowed us to focus some of them to the task of disrupting the activities of the opposing team.

The remainder of the paper is organised as follows. Section~\ref{sec:arch} lists the languages, tools, and IDEs used during the development of our team. In Section~\ref{sec:strategy14} we summarise the main strategies that were imported from our previous participation as team LFC, which is followed by Section~\ref{sec:strategy15} where we describe the new strategies that tackle the main extensions of the scenario. Section~\ref{sec:matches} contains a brief account of our performance in the matches, and in Section~\ref{sec:questions} we present the responses to a questionnaire created by the contest organisers.  Finally, Section~\ref{sec:conclusion} concludes the paper with an overview of our participation in the 15$^{th}$ \acrshort{mapc}.

\section{Languages and Tools}
\label{sec:arch}

\begin{sloppypar}
For our submission to the  15$^{th}$ edition of the MAPC, we used the same set of languages, tools, and IDEs to develop MLFC as when we developed LFC with one exception; an additional tool for formal verification. The agents, environment, and organisation were all developed in the JaCaMo framework for multi-agent oriented programming. The IDE used was Eclipse with the JaCaMo Eclipse plugin, whereas the automated planner used was the \acrfull{fd} planning system.
To facilitate the formal verification of the protocol used for map merging.
The \gls{csp} formal language was used to verify the correctness of the protocol used for map merging; this involved using a new tool --- the \gls{fdr} model checker.
\end{sloppypar}

JaCaMo\footnote{\url{http://jacamo.sourceforge.net/}}~\cite{boissier2020multi,Boissier11} is a framework for developing multi-agent systems based on the multi-agent oriented programming paradigm.
Unlike the more traditional agent-based languages and tools, JaCaMo models both the environment and the organisation as first class abstractions at the same level as agents, through the use of three complementary approaches: \emph{Jason}, \emph{CArtAgO} and \emph{Moise}.
Jason~\cite{Bordini07} is the language used to program the agents based in the \gls{bdi} model~\cite{bratman:87a,rao:95b,rao91a}; CArtAgO~\cite{Ricci09} is used to program the environment in the Java programming language through the representation of artifacts; and Moise~\cite{Hubner07} is used to model the organisation into three different dimensions: structural (roles and groups), functional (goals, missions, and schemes), and normative (norms and obligations). JaCaMo combines these three technologies seamlessly into one integrated framework. 

The base structure of our code remained unchanged from our participation as team LFC in the previous edition. Each agent has its own artifact which acts as a client to communicate with the competition server. Additionally, we have a team artifact that is used as a blackboard to share specific information with all of the agents in our team. Although we are only using the structural dimension of Moise,  we expanded the roles used in MLFC and added automated plans for changing roles without losing track of what the agent was doing in its previous role.

Fast Downward\footnote{\url{http://www.fast-downward.org/}}~\cite{Helmert06,Helmert09} is a well known automated planner that is still being used as a base planning system for many teams that participate in the International Planning Competition\footnote{\url{https://www.icaps-conference.org/competitions/}}. The planner takes as input a domain and a problem specification, written in the Planning Domain Definition Language (PDDL)~\cite{Mcdermott98}, and generates a solution as output. Our representation remains the same as last time: the domain is created at design time and remains static at run-time, whereas the problem is compiled at runtime and includes only those  61 cells that are observable by the agent that invokes the planner.

\gls{csp}~\cite{Hoare1978} is a formal language designed for specifying concurrent communicating systems. A \gls{csp} specification is built out of \textit{processes}, which describe the sequence of \textit{events} that occur as the system evolves. An event is a communication on a \textit{channel} which may declare typed parameters, such that the events occurring on the channel must be composed of parameters matching the channel's type(s). Processes can be offered in sequence, as a choice, or in parallel. Parallel processes may cooperate on a set of channels; they are said to synchronise on the events on these channels, meaning that the events must occur simultaneously in both processes. This is how \gls{csp} processes communicate with each other, but it is important to note that a process that is not in parallel with another may perform events without requiring another process to `receive' them. We used CSP to model our protocol for merging the map information of our agents.

\gls{fdr}~\cite{GibsonRobinson2014} is a model checker for \gls{csp} specifications, which contains various built-in assertions such as deadlock and divergence (livelock) freedom. \gls{fdr} also includes the Probe tool, which allows the user to step through the available events in the specification. We summarise how we used \gls{fdr} to formally verify the map merge protocol in Section~\ref{sec:verification}, and a full description can be found in~\cite{luckcuck_formal_2021}.

\section{Main Strategies taken from the 14th MAPC}
\label{sec:strategy14}

In this section we summarise the three main strategies that we previously used as team LFC in the $14^{th}$ MAPC and that are broadly used (mostly unchanged) in the $15^{th}$ MAPC. These three main strategies comprise: 1) \emph{agent identification}, for identifying other teammates when they are encountered in the grid; 2) \emph{building a map}, how to mutually build and share map information between the team; and 3) \emph{planning}, using an automated planner to perform optimal planning based on the observable environment of the agent.  Other simpler/smaller strategies continue to be used (for example, exploration strategies and their termination conditions); however, they are not relevant to the new strategies discussed in the next section and thus have been omitted here. Full details of all of the strategies used in the $14^{th}$ MAPC can be found in~\cite{Cardoso20d}.

\subsection{Agent Identification}
\label{sec:id}

As agents do not possess information regarding the identity of other agents at the beginning of each round, they first need to identify their teammates in order to cooperate. Thus, as each agent observes another, it can only recognise whether that agent is a member of its own team (i.e., it does not know which agent it is).  One of the first challenges each agent faces, therefore, is one of \emph{agent identification}: when an agent meets another member of its team, it has to try to determine exactly which of its teammates the agent is.

Fortunately, the identification process is not influenced by having a circular map; in fact, the agents only require local information to identify the other members of their team. As all of the information used falls within the agent's observable environment, all coordinates used can be relative to the agent. Thus, it does not matter if some of the positions span across the edges of what would normally be a border. This allows us to reuse the same identification process that we previously used in~\cite{Cardoso20d}. In the following, we summarise the main aspects and features of this process, and then provide a simple example of agent identification in practice.

At the beginning of a step, each agent perceives its environment, through the receipt of a set of \emph{perceptions} from the server. These perceptions are checked by the agent to recognise if there are any observed unknown entities which could be part of its team. If this is the case, a broadcast message is triggered by the agent to ask all of the other agents in its team to communicate the details of what they can currently observe in their respective local environments during that time step.  Each agent that receives such a broadcast message has to reply with a list containing all the \emph{thing} perceptions it has (i.e., the objects seen by the agent). 
This list of \emph{things} is then used by the agent to understand if some agent that it currently observes is the same as that which sent the reply. 

This can be done in two steps.
 First, by determining if the replying agent observes an unknown entity in some position that would correlate with the requesting agent; in which case, the agents may be looking at each other.
To verify this, the second identification step consists of checking if there is also a correlation with all of the other \emph{things} observed by the requesting agent (and their relative positions) with those observed by the responding agent.
If this is the case, it means that not only the two agents are seeing each other, but that all of the \emph{things} in their local environment coincide. This allows the agent to conclude the identification of another team member. Note that this process may generate false positives; for example if multiple agents are in the same exact formation (seeing the same objects, etc.). When this happens, the unknown entity is identified with multiple identifiers (i.e., where one of them is the right one, but it is unclear which of these it is); thus, the identification process fails and the agent is not identified. Because of this, the identification process needs to be constantly reapplied in subsequent simulation steps until all agents are identified.

\tikzset{every picture/.style={line width=0.75pt}} 

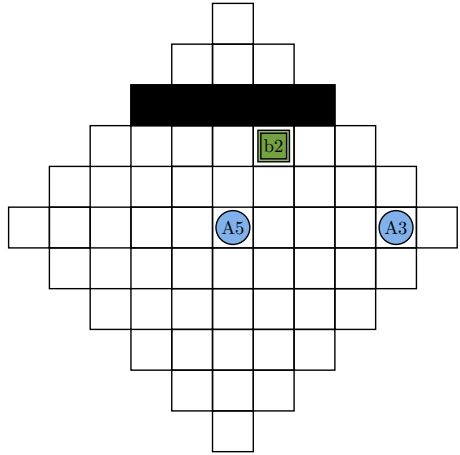
\begin{figure}[ht]
\centering
\scalebox{0.7}{

\begin{tikzpicture}[x=0.75pt,y=0.75pt,yscale=-1,xscale=1]

\draw   (213,214.08) -- (242.38,214.08) -- (242.38,243.47) -- (213,243.47) -- cycle ;
\draw   (213,243.47) -- (242.38,243.47) -- (242.38,272.85) -- (213,272.85) -- cycle ;
\draw   (213,272.85) -- (242.38,272.85) -- (242.38,302.23) -- (213,302.23) -- cycle ;
\draw   (213,302.23) -- (242.38,302.23) -- (242.38,331.62) -- (213,331.62) -- cycle ;
\draw   (213,331.62) -- (242.38,331.62) -- (242.38,361) -- (213,361) -- cycle ;
\draw   (213,184.7) -- (242.38,184.7) -- (242.38,214.08) -- (213,214.08) -- cycle ;
\draw   (183.62,184.7) -- (213,184.7) -- (213,214.08) -- (183.62,214.08) -- cycle ;
\draw   (213,125.93) -- (242.38,125.93) -- (242.38,155.32) -- (213,155.32) -- cycle ;
\draw   (213,96.55) -- (242.38,96.55) -- (242.38,125.93) -- (213,125.93) -- cycle ;
\draw   (213,67.17) -- (242.38,67.17) -- (242.38,96.55) -- (213,96.55) -- cycle ;
\draw   (213,37.78) -- (242.38,37.78) -- (242.38,67.17) -- (213,67.17) -- cycle ;
\draw   (242.38,184.7) -- (271.77,184.7) -- (271.77,214.08) -- (242.38,214.08) -- cycle ;
\draw   (271.77,184.7) -- (301.15,184.7) -- (301.15,214.08) -- (271.77,214.08) -- cycle ;
\draw   (359.92,184.7) -- (389.3,184.7) -- (389.3,214.08) -- (359.92,214.08) -- cycle ;
\draw   (154.23,184.7) -- (183.62,184.7) -- (183.62,214.08) -- (154.23,214.08) -- cycle ;
\draw  [fill={rgb, 255:red, 0; green, 0; blue, 0 }  ,fill opacity=1 ] (242.38,96.55) -- (271.77,96.55) -- (271.77,125.93) -- (242.38,125.93) -- cycle ;
\draw   (95.47,184.7) -- (124.85,184.7) -- (124.85,214.08) -- (95.47,214.08) -- cycle ;
\draw   (66.08,184.7) -- (95.47,184.7) -- (95.47,214.08) -- (66.08,214.08) -- cycle ;
\draw   (183.62,155.32) -- (213,155.32) -- (213,184.7) -- (183.62,184.7) -- cycle ;
\draw   (183.62,125.93) -- (213,125.93) -- (213,155.32) -- (183.62,155.32) -- cycle ;
\draw   (183.62,96.55) -- (213,96.55) -- (213,125.93) -- (183.62,125.93) -- cycle ;
\draw   (183.62,67.17) -- (213,67.17) -- (213,96.55) -- (183.62,96.55) -- cycle ;
\draw   (154.23,155.32) -- (183.62,155.32) -- (183.62,184.7) -- (154.23,184.7) -- cycle ;
\draw  [fill={rgb, 255:red, 0; green, 0; blue, 0 }  ,fill opacity=1 ] (183.62,96.55) -- (213,96.55) -- (213,125.93) -- (183.62,125.93) -- cycle ;
\draw   (154.23,96.55) -- (183.62,96.55) -- (183.62,125.93) -- (154.23,125.93) -- cycle ;
\draw  [fill={rgb, 255:red, 0; green, 0; blue, 0 }  ,fill opacity=1 ] (213,96.55) -- (242.38,96.55) -- (242.38,125.93) -- (213,125.93) -- cycle ;
\draw   (124.85,125.93) -- (154.23,125.93) -- (154.23,155.32) -- (124.85,155.32) -- cycle ;
\draw   (95.47,155.32) -- (124.85,155.32) -- (124.85,184.7) -- (95.47,184.7) -- cycle ;
\draw   (242.38,155.32) -- (271.77,155.32) -- (271.77,184.7) -- (242.38,184.7) -- cycle ;
\draw   (242.38,125.93) -- (271.77,125.93) -- (271.77,155.32) -- (242.38,155.32) -- cycle ;
\draw   (242.38,96.55) -- (271.77,96.55) -- (271.77,125.93) -- (242.38,125.93) -- cycle ;
\draw   (242.38,67.17) -- (271.77,67.17) -- (271.77,96.55) -- (242.38,96.55) -- cycle ;
\draw   (271.77,96.55) -- (301.15,96.55) -- (301.15,125.93) -- (271.77,125.93) -- cycle ;
\draw   (271.77,125.93) -- (301.15,125.93) -- (301.15,155.32) -- (271.77,155.32) -- cycle ;
\draw   (271.77,155.32) -- (301.15,155.32) -- (301.15,184.7) -- (271.77,184.7) -- cycle ;
\draw   (301.15,155.32) -- (330.53,155.32) -- (330.53,184.7) -- (301.15,184.7) -- cycle ;
\draw   (301.15,125.93) -- (330.53,125.93) -- (330.53,155.32) -- (301.15,155.32) -- cycle ;
\draw   (330.53,155.32) -- (359.92,155.32) -- (359.92,184.7) -- (330.53,184.7) -- cycle ;
\draw   (242.38,214.08) -- (271.77,214.08) -- (271.77,243.47) -- (242.38,243.47) -- cycle ;
\draw   (271.77,214.08) -- (301.15,214.08) -- (301.15,243.47) -- (271.77,243.47) -- cycle ;
\draw   (242.38,243.47) -- (271.77,243.47) -- (271.77,272.85) -- (242.38,272.85) -- cycle ;
\draw   (271.77,243.47) -- (301.15,243.47) -- (301.15,272.85) -- (271.77,272.85) -- cycle ;
\draw   (301.15,243.47) -- (330.53,243.47) -- (330.53,272.85) -- (301.15,272.85) -- cycle ;
\draw   (242.38,272.85) -- (271.77,272.85) -- (271.77,302.23) -- (242.38,302.23) -- cycle ;
\draw   (242.38,302.23) -- (271.77,302.23) -- (271.77,331.62) -- (242.38,331.62) -- cycle ;
\draw   (271.77,272.85) -- (301.15,272.85) -- (301.15,302.23) -- (271.77,302.23) -- cycle ;
\draw   (183.62,302.23) -- (213,302.23) -- (213,331.62) -- (183.62,331.62) -- cycle ;
\draw   (183.62,272.85) -- (213,272.85) -- (213,302.23) -- (183.62,302.23) -- cycle ;
\draw   (183.62,243.47) -- (213,243.47) -- (213,272.85) -- (183.62,272.85) -- cycle ;
\draw   (154.23,214.08) -- (183.62,214.08) -- (183.62,243.47) -- (154.23,243.47) -- cycle ;
\draw   (183.62,214.08) -- (213,214.08) -- (213,243.47) -- (183.62,243.47) -- cycle ;
\draw   (95.47,214.08) -- (124.85,214.08) -- (124.85,243.47) -- (95.47,243.47) -- cycle ;
\draw   (124.85,243.47) -- (154.23,243.47) -- (154.23,272.85) -- (124.85,272.85) -- cycle ;
\draw   (154.23,272.85) -- (183.62,272.85) -- (183.62,302.23) -- (154.23,302.23) -- cycle ;
\draw   (301.15,184.7) -- (330.53,184.7) -- (330.53,214.08) -- (301.15,214.08) -- cycle ;
\draw   (330.53,184.7) -- (359.92,184.7) -- (359.92,214.08) -- (330.53,214.08) -- cycle ;
\draw   (301.15,214.08) -- (330.53,214.08) -- (330.53,243.47) -- (301.15,243.47) -- cycle ;
\draw   (330.53,214.08) -- (359.91,214.08) -- (359.91,243.47) -- (330.53,243.47) -- cycle ;
\draw   (124.85,184.7) -- (154.23,184.7) -- (154.23,214.08) -- (124.85,214.08) -- cycle ;
\draw  [fill={rgb, 255:red, 103; green, 152; blue, 46 }  ,fill opacity=0.7 ] (245.71,129.26) -- (268.44,129.26) -- (268.44,151.99) -- (245.71,151.99) -- cycle ;
\draw  [fill={rgb, 255:red, 103; green, 152; blue, 46 }  ,fill opacity=0.7 ] (247.81,131.36) -- (266.34,131.36) -- (266.34,149.89) -- (247.81,149.89) -- cycle ;
\draw  [fill={rgb, 255:red, 74; green, 144; blue, 226 }  ,fill opacity=0.7 ] (333.03,199.39) .. controls (333.03,192.66) and (338.49,187.2) .. (345.22,187.2) .. controls (351.96,187.2) and (357.42,192.66) .. (357.42,199.39) .. controls (357.42,206.12) and (351.96,211.58) .. (345.22,211.58) .. controls (338.49,211.58) and (333.03,206.12) .. (333.03,199.39) -- cycle ;
\draw  [fill={rgb, 255:red, 74; green, 144; blue, 226 }  ,fill opacity=0.7 ] (215.5,199.39) .. controls (215.5,192.66) and (220.96,187.2) .. (227.69,187.2) .. controls (234.42,187.2) and (239.88,192.66) .. (239.88,199.39) .. controls (239.88,206.12) and (234.42,211.58) .. (227.69,211.58) .. controls (220.96,211.58) and (215.5,206.12) .. (215.5,199.39) -- cycle ;
\draw  [fill={rgb, 255:red, 0; green, 0; blue, 0 }  ,fill opacity=1 ] (271.77,96.55) -- (301.15,96.55) -- (301.15,125.93) -- (271.77,125.93) -- cycle ;
\draw  [fill={rgb, 255:red, 0; green, 0; blue, 0 }  ,fill opacity=1 ] (154.23,96.55) -- (183.62,96.55) -- (183.62,125.93) -- (154.23,125.93) -- cycle ;

\draw (257.07,140.63) node   [align=left] {b2};
\draw (345.22,199.39) node   [align=left] {A3};
\draw (227.69,199.39) node   [align=left] {A5};

\end{tikzpicture}

}
\caption{Identification example with two agents $A_5$ and $A_3$, a dispenser of type `$b2$', and black shaded cells representing obstacles.}
\label{fig:identification}
\end{figure}

To better illustrate how the identification process works, we give a simple example 
(Figure~\ref{fig:identification}), where we have two agents on the same team: $A_5$ and $A_3$. 
For the sake of simplicity, we focus on the identification process on $A_5$'s side (for $A_3$ it is symmetric). In this example, we assume that $A_5$ has not identified $A_3$ yet, and when the belief \texttt{thing(4, 0, entity, ``A'')} is added to its belief base, it broadcasts an identification request to each agent in its team. When $A_3$ receives this request, it sends to $A_5$ the list of things it is currently seeing: \texttt{[thing(-4, 0, entity, ``A''), thing(-3, -2, dispenser, b2)]}. With this information, $A_5$ first checks whether there is an entity in the list with coordinates \texttt{(X, Y)}, and a corresponding entity in its belief base with symmetric coordinates \texttt{(-X, -Y)}. In this example, this is satisfied since \texttt{thing(-4, 0, entity, ``A'')} is indeed in the list returned by $A_3$. Moreover, in  $A_5$'s belief base we find \texttt{thing(4, 0, entity, ``A'')} in the corresponding symmetric position. Thus, it is possible that the agent in $A_5$'s local observable environment is in fact $A_3$. In order to be sure that this is the case, $A_5$ needs to check that each entity \texttt{thing(W, Z, Type, Name)} in the list, corresponding to the other observed entities in its local environment defined by ($\mid$\texttt{W+X}$\mid$ \texttt{+} $\mid$\texttt{Z+Y}$\mid$ $\leq$ \texttt{5}), is also present in its belief base. This means we can find \texttt{thing(W+X, Z+Y, Type, Name)} in the belief base for each thing in the list (where \texttt{X} and \texttt{Y} are the relative coordinates of $A_3$ from $A_5$'s viewpoint). 

In this example, we have \texttt{X = 4}, \texttt{Y = 0} and the list containing only one other entity; \texttt{thing(-3, -2, dispenser, b2)}. Since $\mid$\texttt{-3+4}$\mid$ \texttt{+} $\mid$\texttt{-2+0}$\mid$ = $\mid$ 1 $\mid$ \texttt{+} $\mid$ -2 $\mid$ = 3 $\leq$ \texttt{5}, the dispenser should also be in $A_5$'s field of vision, and it is. In fact, we find \texttt{thing(1, -2, dispenser, b2)} in $A_5$'s belief base. Assuming that there are no other teammates at the same distance, $A_5$ can safely conclude that the agent in its field of vision is $A_3$. The same process is done by $A_3$, which will identify $A_5$ as the agent in its own field of vision.

\subsection{Building a Map}
\label{sec:map}

In order to move purposefully within the map and to perform tasks
efficiently, agents dynamically build a map storing information of the
environment that they observe. The map is built following an approach
based on the one used in the previous year's scenario~\cite{Cardoso20d}, where
each agent considers its starting position as the $(0,0)$
coordinate of its local map, and stores in this map all the relevant
information collected while moving and perceiving entities 
it observes in its local environment. Thus, the coordinate system of
each map will be relative to the starting position of each agent.
Functionally, these maps are stored in a CArtAgO artifact 
as \emph{HashMaps} where the key is the name of the owner of the 
map and the elements are sets of points representing cells. Furthermore, 
local maps are merged when meeting other agents.

The main challenges that needed to be addressed in migrating our agents from the
14$^{th}$ to the 15$^{th}$ \acrshort{mapc} are:
(1) do the agents
need to store more (or conversely less) information regarding entities encountered in the environment, than in the previous year's scenario; and
(2) how should the new borderless and continuous map be managed.

The environment consists of a number of entities, some of which are immutable (i.e., the location of the entities remains static throughout a full round), and some of which are dynamic.  Whilst some entities are highly dynamic (for example, the agents themselves), other entities can be affected by agent behaviour, such as the movement of a block or the result of a \emph{clear} event\footnote{Clear events occur randomly, but agents have access to a clear action which has a reduced area of effect but otherwise functions the same. A clear event/action will remove any obstacles or blocks and disable any agents that are inside its area of effect.}.  Thus, as the agent moves, it observes its local environment at each time-step, and updates its local map with the location of the following static entities: \emph{dispensers}, \emph{goal positions} and \emph{taskboards}.  Whilst there could be utility in retaining information about dynamic entities such as the position of obstacles, blocks and agents, this information may become stale over time and thus is unreliable.  Planning (Section~\ref{sec:smartmove}) therefore only takes into account knowledge of previously observed static entities, and currently observed entities within the local environment.

Due to the fact that the grid map in this version of the scenario differs from the previous one in that the map is now circular, the map that we store with our previous strategy could appear infinite in scale, with each entity appearing multiple times with different coordinates (i.e., multiples of the actual size of the map itself).  Thus there is the risk that an agent may perceive multiple instances of the same entity.  To address this problem the cartographers compute and broadcast the size of the map
(see Section~\ref{sec:carto} for details). Once the information about
the size of the map is received by the agents, the maps are normalised
accordingly, thus facilitating the easy identification of
potential repetitions. The normalisation of the map is not only
important to avoid repetitions of elements, but it also allows
agents to establish the shortest distance to potential points of
interest (e.g., dispenser or goal positions) by going around (of what would normally be) the edges of the map.

The discussion so far has focused on how local maps are built, but
the collected information also need to be shared when agents from the same team meet each other. This is when the process of merging maps
happens. To start with, all local maps are already stored in the
\gls{team:teamart}, meaning that in theory each agent could have
access to the other agents' maps. This, however, would not be beneficial
if the maps are not merged, as the coordinate system of each agent
differs. The full details of the merging process is described 
in~\cite{Cardoso20d}, but it is based on the following intuition: suppose
that there are two distinct groups of agents $G_1$ and $G_2$, where all agents in $G_1$ have already
merged their maps, and likewise, all agents in $G_2$ have also merged their
maps. The map merging process starts when an agent $A_1$ from $G_1$
meets an agent $A_2$ from $G_2$. The two involved agents, $A_1$ and
$A_2$, communicate all the relevant information (e.g., their positions
and the position of the agent they are seeing) to the respective
leaders of the groups; let us call them $L_1$ and $L_2$, and the
merging process is delegated completely to the leaders. The leaders 
decide who is going to be the new leader of the group $G_1\cup G_2$,
and merge the information by taking care that no repetition occurs and
that all the agents in $G_1 \cup G_2$ share the same coordinate system.

\subsection{Planning}
\label{sec:smartmove}

As mentioned in the previous section, the dynamic nature of the map
and agents moving blocks can make the planning of a route challenging.  For
this reason, our agents plan their next few actions based on
what is observable in their current environment and the static entities stored in the merged map. We use \emph{task planning} as opposed to path planning for the movement of the agents, as we also need to consider other actions (such as the \emph{clear} action) that can alter the way our agents move. Furthermore, since we are dealing with a 2D grid, the problem is simplified as the cells can be modelled as states, and there is no need to use advanced path planning algorithms (due to the fact that we do not have live noisy sensors, nor do we care about collisions).

Planning is not utilised during the initial phase, whereby agents explore the map and compute the map's size.  However, once the agents start
collaborating with each other, they have sufficient information for planning tasks regarding the existence and location of dispensers, goal positions and taskboards, based on destination information retrieved from the merged maps. 
The process of reaching the desired destination is an
iterative process composed of three stages: 1) selection of a proximal destination that falls within the agents local, observable environment (en route to the actual destination); 2) invocation of the
\acrshort{fd} planner; and 3) managing the planner's result.

When an agent has a task assigned to it (e.g., heading towards a
taskboard or dispenser), it retrieves the global coordinates of the
closest cell satisfying its requirements, and translates them into
relative coordinates. Then, the agents selects a \emph{good cell}
within its observable environment which minimises the overall distance to the actual destination.  Such a cell is considered good provided that 
it does not contain any agent or block
(more details on the procedure for the selection of a good cell can be
found in~\cite{Cardoso20d}). If no good cell can be found, then the agent behaves as if the \acrshort{fd} planner has returned an empty plan,
which is described later in this section.

Once the agent has selected a good cell, it calls the \acrshort{fd}
planner via the \gls{team:agentart}. In order to invoke the planner,
several pieces of information need to be provided: what elements are
currently observable by the agent, and how they are represented; what actions are
allowed; the goal itself; and whether or not the agent has a block
attached to it. First, the details of the agent's observable local environment are already present in the \gls{team:agentart}. This allows for the
creation of a problem file composed of 61 cells (all the currently observable cells) where the agent is in the centre, dispensers and taskboards
are not represented (i.e., they do not constitute obstacles to
movements), obstacles are considered obstacles, and blocks and other
agents are modelled as blocks. The latter is due to a conservative
approach where the planner is not allowed to clear any block, and it is hard to know on-the-fly whether or not the perceived blocks belong to the enemy team (furthermore, this prevents the possibility of us clearing our own blocks by mistake). Second, movement and rotation actions are always permitted, but the clear action is allowed only if the agent has enough energy. From a modelling point of view, allowing or not allowing
  a clear action is based simply on using a PDDL domain file with or
  without the definition of the action. Finally, the remaining
information (i.e., what the goal is, and whether the agent has a block
attached) is provided to the \gls{team:agentart} by the agent. Once
the \gls{team:agentart} has collected all the required information, it
produces the problem\footnote{An example of a problem file that was generated dynamically during one of the matches can be found at: \url{https://github.com/autonomy-and-verification-uol/mapc2020-lfc/blob/master/planner/example_problem.pddl}} and domain files used by the \acrshort{fd}
planner; the planner is then invoked and the results are collected.

Any returned solution (empty or not) is processed by the agent. If a
sequence of actions is returned, then the agent blindly executes them
in the same order as that within the solution, despite the fact that this
can result in some of the
actions failing. In this scenario, we adopt a forgiving approach whereby
the failure of an action does not necessarily imply that the planner should be
called again.\footnote{Failures of a movement action are tracked for
  the agent to have an up-to-date idea of the distance to its
destination.} The rationale here is that, provided that not all of the actions fail, the agent will still approach the destination, and it
reduces the amount of required resources.

The situation is different when the returned plan is empty, or if the
agent was unable to locate a good cell in its local environment. In this case,
the agent is required to make its own decision regarding its next
action.  Our approach is for the agent to look for a one-step movement
action, which brings the agent closer to its destination, and then to
re-invoke the planner. The idea here is that given the dynamic nature of the
environment, by making a move towards the destination, knowledge of
the local environment will improve, possibly resulting in the 
generation of
a valid plan. Such a heuristic has so far resulted in  good performance of our
system, but it is clearly not optimal yet.

It is easy to see how the whole process of planning via the use of
the external \acrshort{fd} planner does not scale well with an
increasing number of agents. To address this issue, our new strategy employs the use of a plan cache, which is described in
Section~\ref{sec:cache}.

\section{New Strategies for the 15th MAPC}
\label{sec:strategy15}

In this section we describe the details of the main strategies that were developed specifically for the 15$^{th}$ MAPC. These include: \emph{cartography}, to discover the size of the map; \emph{verification of map merging}, to provide assurances about the reliability of the map merging protocol; \emph{plan cache}, to allow up to 50 agents to use the planners efficiently; \emph{bullies}, these are agents that focus solely on disrupting the opposing team by trying to clear their blocks; and \emph{achieving tasks}, to use the new taskboard facility and decentralise task assembly into multiple groups.

\subsection{Cartography}
\label{sec:carto}

One of the main objectives of our implementation is to maintain a map that
is common across all of our agents that have identified each other at some
point in the simulation. In the 14$^{th}$ MAPC, we were able to identify
suitable goal positions and block dispensers by their relative
position to the borders of the map. Furthermore, in the initial exploration phase agents randomly chose a direction to explore, and only changed this direction when they approached the border of the map. Since the scenario of this year used a spherical or \emph{circular} map, where the map ``wraps-around'' on the sides, we
can no longer rely on the existence of borders. 

Having explored the 
underlying issues, we realised that the main information we were missing
to re-use the previous strategies was the \emph{size} of the map.
To obtain the size of the map, we introduced a new role for agents: \emph{cartographers}. The aim of this new role is to find the
exact size of the map, in both horizontal and vertical directions (which can be different, for example, we can have grids that are 60x50).
The size determination occurs in a preliminary phase, before the agents
can start with the identification of suitable building sites and assembling blocks for tasks. Cartography is always approached by two agents at once for
either horizontal or vertical direction. It is necessary that these two agents have identified each other, and in fact this is what triggers their desire to become cartographers. Only two pairs of cartographers are
required; one pair for each dimension (i.e., vertical and horizontal). As soon as two
agents adopt the cartographer role for one dimension, they will start to move into opposite directions along that dimension, counting the successful steps they make, until they meet again. For example, if \(A_1\)
and \(A_2\) start to work as a pair of cartographers for discovering the size of
the horizontal axis, one of them, say \(A_1\) will start
to move to the left, while the other, \(A_2\) will move to
the right. Since the map folds onto itself, they will necessarily
meet again, and can add the number of steps they took to compute
the size of the map along that horizontal dimension\footnote{When \(A_1\)
and \(A_2\) first adopt the role of cartographers, they need to retain the distance between them, as this initial distance has to be added to the final sum.}. This information is then broadcast to all of the other agents in the team. 

The cartographers also need to be able to maintain their movement along the axis that they are responsible for exploring. Blocks and obstacles can be cleared by using the clear action. At this stage we do not have to worry about clearing our own blocks, since our other agents are tasked with exploring the grid and will not collect any blocks until the cartography phase has concluded. The only remaining impediment in the path of our cartographers are other agents (friendly and enemy). Friendly exploration agents already avoid other agents they meet on their path by moving around them (e.g., if the are moving south and encounter another agent, they will shift to either east or west, and then continue moving south for a few steps before realigning themselves). We initially planned a similar behaviour for our cartographers in order to avoid enemy agents, tracking the number of cells that the agent moved to ensure that the final calculation remained correct, but due to time constraints this implementation was not finalised in time for the contest. Instead, during the contest if our cartographers were to meet enemy agents along their path that would block their movements, they would simply keep trying to move in the same direction until successful.


Two exploring 
agents will assume the cartographer role under the following conditions:
\begin{itemize}
    \item  They have just identified each other (as described in Sect. \ref{sec:id}).
    \item The size of the map in at least one dimension is unknown by the team.
    \item There is at least one dimension that has no assigned pair of cartographers that are actively determining its size.
\end{itemize}

The first condition ensures that the agents can identify each other and thus determine when they meet again having
covered the whole length of the map.  The other conditions ensure that we have exactly one pair 
of cartographers for each dimension, while allowing all other agents focus on either exploring the map or disrupting
the opposing team (see Section~\ref{sec:bully} for the latter). Due to the nature of the maps in the 15$^{th}$ MAPC,
agents often start in clusters, which means that the assignment of the four cartographer roles typically occurs during
the first step of the simulation.

\subsection{Formal Verification of Map Merging}
\label{sec:verification}

The map merging protocol -- which controls how to merge individual agent's maps -- was built for the 14$^{th}$ MAPC, and our team reused it for this new edition of the contest.
Section~\ref{sec:map} summarises the protocol, which is described in detail in~\cite{Cardoso20d}. In brief, the protocol allows a map used by a group (one or more) agents to be merged into another map, pooling the information in both maps and unifying the coordinate system, which enables the group to cooperate. This involves message-passing between several agents (at least 2 and at most 4 per instance of the merge protocol) to coordinate the merge (messages to more agents can be sent after the merge has been completed to update their information). 

Having a unified map is critical to the team being able to assemble and deliver tasks, which is why we chose to apply formal verification to the map merging protocol. The protocol had worked well in the 14$^{th}$ MAPC, but testing the protocol was difficult because of the changing environment and interference from the other team in the match. 

Verification by Formal Methods encompasses a
wide range of mathematically-defined techniques for describing how a system should behave or how it will operate on data, and tools for reasoning about the correctness of these specifications. As mentioned in Section~\ref{sec:arch}, we specified the map merging protocol in the formal language \gls{csp} and used the model checker \gls{fdr} for both validation and verification, as described below. A complete description of the specification and verification effort is presented in~\cite{luckcuck_formal_2021}. 

The specification was built manually, from careful examination of the agent plans and the description of the protocol in~\cite{Cardoso20d}. The specification kept close architectural correspondence with the agent program and protocol: each agent was represented by its own process and given an agent ID, and the agent plans involved in the merge process were represented by events of the same name. This helped with tracing parts of the specification back to their source in the program. The formal specification made some abstractions from the implemented protocol, for example we did not model the agent reasoning cycle.  Furthermore, the specification contained three agents; the merge protocol happens between two agents, and the third lets us check that the protocol performs correctly when there is interference.

The protocol specification was also \emph{validated} and \emph{verified}.
Borrowing Boehm's descriptions~\cite{boehm_verifying_1984}, \emph{validation} answers the question ``Am I building the right product?'' whereas \emph{verification} answers the question ``Am I building the product right?''. Validating our specification involved checking that it matched the behaviour of the implementation. To verify our specification we checked that it was itself correct and could perform the protocol's behaviour without errors. 

Simple validation checks were performed using \gls{fdr}'s Probe tool, which allows a user to step-through the specification. This was invaluable for specification debugging. For more complicated and repeatable checks, we used \gls{fdr}'s built-in $[has~trace]$ assertion, which checks that the specification can perform a given trace of events without divergence (livelock), without refusing any of the events in the trace. This assertion was used to check the specification's behaviour in six different scenarios, which were based on the protocol implementation's behaviour during the 14$^{th}$ MAPC. The $[has~trace]$ checks were performed automatically by \gls{fdr}, which made them easy to rerun after updates to the specification (similar to performing regression testing). 

For verification of the specification, we checked that it was free from divergence and non-determinism (where the specification may perform several different events, after a given prefix), and that the specification could reach a \textit{done} state --- where a $done$ event occurs when there is only one map, shared by all the agents. The divergence and non-determinism checks use \gls{fdr}'s built-in assertions, but the \textit{done} check was added to the specification manually.
Reaching the \textit{done} state shows that the protocol has behaved correctly and implies that the specification did not deadlock along the way. 

We found that \gls{csp} was well suited to specifying the map merging protocol, because its features focus on communication and concurrency. \gls{fdr} provides features that are helpful during specification debugging and verification. The verification effort helped provide additional confidence that the merge protocol worked correctly, and showed how \gls{csp} could be applied to \gls{mas} communication protocols. In our previous work we applied \gls{csp} to one module of a single autonomous robotic system, as part of a suite of formal verification approaches~\cite{cardoso_heterogeneous_2020,cardoso_towards_2020}. This verification work shows the utility of applying \gls{csp} (and formal verification approaches more generally) to aspects of \gls{mas} as well.

\subsection{Plan Cache}
\label{sec:cache}

Using automated planning at runtime is far from being an easy task. The entire process, from encoding the state of the agent as a problem file, to solving it with a planner, is computationally demanding. This can be reasonable for small/simple applications, but can become an issue when applied to large and complex systems, such as a MAS. Indeed, even though the planning problem for a single agent is feasible, it might not be for a coalition of agents where each agent has to perform its own planning as well.
Assuming that we had $N$ agents, this would require us to call the planner $N$ times (one per agent).
Furthermore, given the MAPC scenario, this could happen for every step of the simulation. Even though individually the problems to be solved are relatively small (remember that we only consider the 61 cells that an agent can observe), calling up to 50 instances of planners (e.g., assuming every agent needs to call the planner in the same step) consumes too many resources. This may not have been a problem if the only thing an agent was required to do in a step was to plan its movement, but in reality there are many operations that every single agent needs to do each step (updating belief base with the dozens of perceptions coming from the server, communicating with other agents, etc.). Of course, more computation power could have also solved (or alleviated) the problem, but we were limited to using laptops (with powerful specifications, but not at the level of high performance computing).

Since physical resources (CPU, memory) and time (how long an agent can wait) are finite, it is always possible to pick a number $N$ of agents for which it is not possible to solve the planning problem in less than a certain amount of time (4 seconds for each simulation step in the contest). In particular, as we mentioned previously, the 15$^{th}$ edition of the contest extended the scenario to have 15 agents in round 1, 30 agents in round 2, and 50 agents in round 3. Through testing, we noticed that our previous strategy managed to hold up for 15 agents, but it did not work for 30 and 50 agents. Because of this, alternatives to speed up the planning process had to be considered.

We investigated alternatives to speed up the process, and found out that we could make the planning process faster by \emph{caching} the plans. By caching, we refer to the act of storing previously generated plans, instead of simply executing and then forgetting about them. When an agent asks the planner to solve a problem, if such a problem has been already solved in the past, it would be more efficient to retrieve the solution found for this problem, than to generate it again using the planner, resulting in redundant work.
This can be achieved by keeping a mapping between $Problem \rightarrow Plan$, which given a problem file, returns its corresponding plan (if present in the cache). When a problem file does not find a match in the cache, it means that it has never been solved before (i.e., a cache miss), in which case the execution continues by calling the planner and then updating the cache. Note that this cache is saved independently of the size of the grid, number of agents, or any other parameter. This means that the cache can be used for any configuration to speed up the planning process.

The first aspect we have to consider for plan caching is how to encode a problem file, so that its retrieval from the cache can be straightforward. Such an encoding must uniquely identify a problem file. Thus, all the information which characterises the agent's local environment needs to be considered.

Given details of an agents local environment,
a possible numerical encoding can be obtained by unrolling the grid as a one-dimensional array.  This unrolling 
starts from the upper most cell, and then all rows are appended one by one from left to right. The encoding works as follows: empty spaces are mapped\footnote{The mapped values are not semantically relevant, as long as is preserved for all mappings.} to 0 (dispensers and the agent current position are considered empty), obstacles to 1, blocks to 2, and the movement target (plan goal) to 3. After the contest we realised that since blocks are considered as obstacles for the planner, we could have set their value to 1, which would have decreased the number of cached plans by at least a small margin.

Once the encoding has been created, it can be used to query the cache. Since we want to reuse the cached plans across different executions, the cache is stored in the secondary memory;
specifically, each cached plan is stored in a separate file, where
the encoding of the agent's local environment is used to name such a file. 
Consequently, to check if a certain planning problem has already been solved, it is sufficient to check to see if 
a file exists that is named as the encoding of the problem. If such a file exists, there is no need to call the planner since the corresponding plan is already available within the file; otherwise, the planner is called and a new file is stored (named after the encoding of the problem).

As we have seen previously, it is possible to call the planner in two different settings. 
The first one corresponds to the generation of a plan by the agent itself; the encoding of which is discussed above.
The second occurs when the agent possesses a block (i.e., a block is currently attached to that agent), and requires a plan.
The encoding for this is almost identical, except that in this case we append the local environment coordinates of the block that is attached (note that our planning domain only supports movement with no more than a single attached block). For example, if there is a block attached to the agent and the block is located one cell below the agent, then we would append the string `01' to the beginning of the encoding.

\subsection{Bullies}
\label{sec:bully}

For the 15$^{th}$ MAPC we decided to add attack strategies for our agents. We created a new role for this called \emph{bully}, whereby a bully is an agent with a single and specific purpose: to clear the blocks used by the enemy team. In our solution, two typologies of bullies were used: \emph{bouncer bullies} and \emph{hunter bullies}.
One of the disadvantages of the strategies adopted by our team for the 15$^{th}$ MAPC was that time was spent during the initial phase to determine the dimensions of the map, prior to building blocks and achieving tasks, whereas other, faster teams could use this time to complete the tasks, thus gaining an advantage.    
To counter this, we developed \emph{bouncer bullies}, which had the task of slowing down the enemy team in the initial phase of the match (i.e., until our agents were ready to start building tasks), by disrupting their ability to complete and submit tasks.
To achieve this, when a bouncer bully finds a goal position, instead of moving away as normally the explorers would do, it starts patrolling such a goal position and the ones close to it. Every time the bully sees an enemy agent with a block, it tries to clear the block. Since clearing requires three steps, the bully does not always succeed. Nonetheless, by being close to a goal position it puts itself in an advantageous position with respect to any approaching enemy agent; to deliver the task the enemy agent has to approach the bully, which increases the possibility of a clear action succeeding.

Once our agents have finished the initial phase and can commence task building, we no longer require bouncer bullies,
as agents with the bouncer bully role could be better employed performing another task, rather than staying at a single goal position.  For example, it might be possible that a bully agent selected a goal position which is never used by the enemy team.
If that is the case, then the bully agent would just waste time patrolling endlessly a goal position where no blocks would ever be cleared. Because of this, in the remaining steps of each simulation (after the cartography phase has concluded), we need a different kind of bully, the \emph{hunter bully}. These agents behave similarly to the bouncer ones; however a key difference is that they move amongst different goal positions.  Specifically, a hunter bully starts patrolling a goal position for a finite number of steps, after which, if no enemy agent with a block was observed, it moves to a different goal position (usually in a different goal cluster). In this way, if the agent initially picked a bad goal position, it will eventually arrive in a good one (i.e., one occupied by the enemy team).

For both bouncer and hunter bullies, the policy to defend a goal position is the same. The agent moves in a circle around the goal position (similar to a shark behaviour). In this way, with respect to staying put, the agent has more possibilities to intercept an enemy agent and clear its blocks.

\subsection{Achieving Tasks}
\label{sec:tasks}

Our strategy for assembling block structures and achieving tasks remains similar in principle to our previous participation~\cite{Cardoso20d}. Before we explain our strategy, we give a couple of reminders about how the scenario works in regards to tasks: to deliver a task, an agent must be inside a goal cell; it is always the case that block structures must be delivered from the position below the agent, i.e., blocks need to be attached south of the agent for them to be delivered.

The previous strategy included the following agent roles: 
\begin{itemize}
    \item \emph{origin agent}: moves to a unoccupied goal position in the bottom-most cell of a goal cluster;
    \item \emph{retriever}: goes to a dispenser, obtains one block of a type of block required, and moves to one of the available positions around the origin agent.
\end{itemize} 
Once the exploration phase ends, our agents would form one group consisting of 1 origin agent and 9 retrievers (in the previous edition of the MAPC all three rounds had 10 agents).

Due to the inclusion of \emph{taskboards} in the 15$^{th}$ MAPC, we created a new role called \emph{deliverer}. The deliverer waits next to the closest taskboard in relation to the expected goal position of the origin agent. Once enough agents are in place, an appropriate task will be accepted by the deliverer, and retrievers will start moving towards the origin agent to build the block structure required by the accepted task. At the same time, the deliverer will make its way to the position on top of the origin agent, or if that is not possible then any position around the origin that is not below it. Once the building phase is complete, the origin will detach from the block structure and change places with the deliverer, which will then attach to the block structure and deliver the task and will become the new origin. The first retriever to bring its block to the origin will become the new deliverer and will move to the taskboard. This allows us to speed up the time spent between tasks. The previous origin will then become a retriever and go to fetch a new block from a dispenser. An example of such configuration is shown in Figure~\ref{fig:complete_example}.

\input{complete_example_fig}

Finally, to make use of the increase in the number of agents, we divided the team into multiple groups. Since each round varies the number of agents (15, 30, and 50), we calculated the number of groups based on the current size of the team in the round: $GroupSize = RoundSize \div 15$. Thus, for rounds with 15 agents we have one group, for 30 we have two groups, and for 50 we have 3 groups (with 5 remaining agents as leftovers). Since we have the addition of two new roles (deliverer and bullies), we also had to plan how many of these roles would be available within a group. A group of 15 agents (agents can join a group later, and in doing so follow this list of priority) has 1 origin agent, 1 deliverer, 12 retrievers, and 1 bully. In the third round (i.e., 50 agents), the last 5 remaining agents will become bullies that are not affiliated with any group.

\section{Match Analysis}
\label{sec:matches}

In this section we provide a brief summary of our performance in the 15$^{th}$ MAPC. We divide our matches into two groups: the first corresponds to the matches played in the first day of the competition, and the second to the matches played in the second day. Each of the five teams had two matches per day.

During the first day, our solution did not perform very well, out of 6 rounds we had 1 win, 1 tie, and 4 losses. This was even more evident in the scenarios with more agents (teams of size 30 and 50, rounds 2 and 3 resp.). The reason for this was due to the computation power required to run the 30 (resp. 50) agents on our machine. Since our solution requires a lot of coordination amongst the agents, when one agent died, it caused a knock-on effect on the other agents, forcing us to restart the whole team. When many agents had to be handled by the machine, sometimes it happened that not all the computations were finished before the deadline (4 seconds). Specifically, this would happen when a lot of agents were calling the planner, which is the most time demanding component of our solution. As this happened many times during the first day, our solution performed poorly and we lost most of the rounds. For instance, many times our agents were almost ready to start submitting tasks when one of the agents died (lost synchrony with the steps from the server), which caused us to have to restart the whole team. Consequently, restarting the team meant starting from scratch, including doing all of the cartography and exploration again. The overall score obtained by each team in the first day is shown in Table~\ref{tab:day1}.

\begin{table}[ht]
	\centering
	\caption{Total score of each team for the first day, with each round being the sum of the two matches that happened in that day.  Our team is shown in {\bf bold}.}
	\begin{tabular} {l@{\hspace{2em}} c@{\hspace{2em}} c@{\hspace{2em}} c@{\hspace{2em}} c@{\hspace{2em}}} \toprule
		Team & Round 1 & Round 2 & Round 3 & Total Score\\ \midrule
		\gls{fit}         & 38 & 166 & 188    & 392 \\
		{\bf \acrshort{mlfc}}       & {\bf 74} & {\bf 44} & {\bf 66}        & 	{\bf 184}  \\
		\gls{dtu}         & 252 & 100 & 105     & 457  \\ 
		\gls{lti}         & 16 & 34 & 56     & 106 \\
		\gls{jac}         & 0 & 18 & 10     & 28 \\
		\bottomrule
	\end{tabular}    
	\label{tab:day1}
\end{table}

During the second day, our solution performed much better, achieving 6 wins out of the 6 rounds, despite the fact that
no changes had been made in the code of our solution between the two days.  The reason for this improvement in performance is due to the 
planner's cache. During the first day of the competition, we found that the application did not have enough data from previous matches to build an efficient plan cache; particularly for the two rounds where there were larger numbers of agents (30 and 50 agents respectively), as most of the tests performed prior to the competition were conducted using only 15 agents.
Consequently, almost all of the calls to the planner resulted in a cache miss and had to be fully evaluated. Since the planner is the most time demanding component of our solution, by having to compute too many plans, the risk of taking more than 4 seconds was high (and if this happened, the agent would lag behind and eventually become useless). Because of this, our solution performed poorly in the first day. Nonetheless, we recovered in the second day by having a richer cache of previously computed plans, and our solution outmatched most of the other teams. By having a richer cache, most of the requests for a plan could be satisfied without actually calling the planner. In this way, the computations were lighter, and the risk for an agent to fall behind reduced considerably, which in turn meant we had to restart our team much less often, allowing tasks to be completed successfully.

\begin{table}[ht]
	\centering
	\caption{Total score of each team for the second day, with each round being the sum of the two matches that happened in that day. Our team is shown in {\bf bold}.}
	\begin{tabular} {l@{\hspace{2em}} c@{\hspace{2em}} c@{\hspace{2em}} c@{\hspace{2em}} c@{\hspace{2em}}} \toprule
		Team & Round 1 & Round 2 & Round 3 & Total Score\\ \midrule
		\gls{fit}          & 247 & 252 & 596   & 1095 \\ 
		{\bf \acrshort{mlfc}}       & {\bf 220} & {\bf 221} & {\bf 383}        & 	{\bf 824}  \\
		\gls{dtu}          & 534 & 235 & 0    & 769  \\ 
		\gls{lti}          & 28 & 48 & 8    & 84 \\
		\gls{jac}          & 20 & 22 & 18    & 60 \\
		\bottomrule
	\end{tabular}    
	\label{tab:day2}
\end{table}

Except for the reasons reported above, the performance of our solution has not been influenced greatly by the presence of an enemy team. The only case where our solution had some issues was when the enemy team used agents to clear our blocks. For instance, there was a match where an enemy agent was patrolling a goal cluster and was clearing all the blocks in its local environment. In such scenario, our origin agent failed to build and deliver the tasks because of this. Another peculiar situation happened in a match where an enemy agent was moving around one of our origin agents and managed to steal the block structure that we were building. This happened because at some point our origin agent detaches from the structure to switch places with the deliverer (the agent that accepted the task in the taskboard and thus must be the one to deliver the task). At this exact moment the enemy agent managed to attach to the structure, whereas normally it is not possible to attach to a block if there is any agent from the other team connected to it or to any other block that is connected to the first block.

\section{Team Overview: Short Answers}
\label{sec:questions}
  \subsection{Participants and their background}
  \begin{description}
    \item \q{What was your motivation to participate in the contest?}
    \\ 
    Three of our members participated in the $14^{th}$ Multi-Agent Programming Contest as team LFC. We decided to participate in the $15^{th}$ MAPC because the scenario was the same (with some extensions), so we could use most of our existing code and then focus on improving it.
    
    \item \q{What is the history of your group? (course project, thesis, $\ldots$)}\\
     We had three new members for the $15^{th}$ edition of the contest. Even though at that time most of us were postdoctoral researchers at the University of Liverpool, many of us have changed affiliations since then, but decided to continue to collaborate and participate in the latest MAPC.
    
    \item \q{What is your field of research? Which work therein is related?}\\
    Our members have worked in many different areas of research, but at the moment the intersection of the knowledge in our group relates to formal verification and logical reasoning. Some of our members also have a strong background on agent programming and agent-based tools, while for others they were aware of it but did not have much experience. 
    
  \end{description}

  \subsection{Statistics}
  \begin{description}
    \item \q{Did you start your agent team from scratch or did you build on your own or someone else's agents (e.g. from last year)?}\\
    Our code is based on our participation (team LFC) in the $14^{th}$ MAPC.
    
    \item \q{How much time did you invest in the contest (for programming, organizing your group, other)?}\\
    Approximately 200 hours.
    
    \item \q{How was the time (roughly) distributed over the months before the contest?}\\
    Some of the months we were able to dedicate more time to it, but there was a spike in activity in the month before the qualification and then another one in the month before the contest.
    
    \item \q{How many lines of code did you produce for your final agent team?}\\
    A total of 9,805 lines of code, with 3,570 lines in Java (including test files and environment artifacts) and 6,235 lines in Jason agent code.
    
    \item \q{How many people were involved?}\\
    Our team had six members.
    
    \item \q{When did you start working on your agents?}\\
    We started working on April $7^{th}$ 2020, but implementation was limited to a couple of days during some of the months that followed.
  \end{description}

  \subsection{Technology \& Techniques}
  \textbf{Did you make use of agent technology/AOSE methods or tools? What were your experiences?}
  \begin{description}
    \item \q{Agent programming languages and/or frameworks?}\\
     We used the Eclipse IDE with the JaCaMo plugin. Some of our members were already familiar with it and it has performed very well in past contests.
     
    \item \q{Methodologies (e.g. Prometheus)?}\\
    We did not use any AOSE method.
    
    \item \q{Notation (e.g. Agent UML)?}\\
    We used normal UML sequence diagrams for specifying some of the protocols we created, but without any agent notation. The sequence diagrams were sufficient to specify what we needed.
    
    \item \q{Coordination mechanisms (e.g. protocols, games, \dots)?}\\
    Moise (part of JaCaMo) was used for the coordination of agents, especially for task coordination. Some of our communication protocols were implemented ad-hoc just using message passing in Jason (based on the sequence diagrams).
    
    \item \q{Other (methods/concepts/tools)?}\\
    We used the Fast-Downward planner for performing efficient AI task planning. The planner was used off-the-shelf with no modifications required.
    
  \end{description}

  \subsection{Agent system details}
  \begin{description}
    \item \q{How do your agents decide \textbf{what} to do?}\\
    Our agents evaluate the beliefs coming from the server at any given step, and based on this information they decide what is the best course of action. Their decision can change in the middle of a step from incoming messages of other agents in the team (e.g., requesting for help).
    
    \item \q{How do your agents decide \textbf{how} to do it?}\\
    Most actions are straightforward, but for long distance movement we call an AI task planner that will plan the best route for the agent (it does so iteratively, based on the observations of the agent on the local environment).
    
    \item \q{How does the team work together? (i.e. coordination, information sharing, ...) How decentralised is your approach?}\\
    Relevant team information is shared on a blackboard (CArtAgO team artifact) to save communication time, but for the most part everything is decentralised.
    
    \item \q{Do your agents make use of the following features: Planning, Learning, Organisations, Norms? If so, please elaborate briefly.}\\
    Our agents use an AI task planner for planning their movement for any phase that comes after the exploration phase. A Moise organisation is used to help to coordinate the agents during task assembly and delivery.

    \item \q{Can your agents change their general behavior during runtime? If so, what triggers the changes?}\\
    Our agents can be explorers, cartographers, deliverers, task origins, retrievers, and bullies. All agents start as explorers but upon meeting certain conditions they will swap to another role/behaviour.
    
    \item \q{Did you have to make changes to the team (e.g. fix critical bugs) during the contest?}\\
    No, some small changes were attempted to improve the amount of no actions being sent to the server, but they were unsuccessful. 
    
    \item \q{How did you go about debugging your system? What kinds of measures could improve your debugging experience?}\\
    Due to the use of several separate tools, debugging the system proved to be quite hard. More time spent in debugging rather than implementing new ideas may provide better results.
    
    \item \q{During the contest you were not allowed to watch the matches. How did you understand what your team of agents was doing?}\\
    We generated some useful logs that could give us an indication of what was happening, however this should be improved in future versions of our team as the logs used were not very intuitive.
    
    \item \q{Did you invest time in making your agents more robust/fault-tolerant? How?} \\
    We have formally verified the protocol we made to merge map information~\cite{luckcuck_formal_2021}. Other than this we did not invest much time apart from the last few days before the contest when we tried to make the team more fault-tolerant towards task failures.
  \end{description}

  \subsection{Scenario and Strategy}
  \begin{description}
    \item \q{What is the main strategy of your agent team?}\\
    Three main new strategies were developed for this edition of the contest: (a) a cartography system that can scout the map and determine the size of the grid; (b) using cached plans that are automatically generated by an AI planner so that our agents can efficiently move through the map; and (c) bully agents that can disrupt the enemy team.
    
    \item \q{Please explain whether you think you came up with a good strategy or you rather enabled your agents to find the best strategy.} \\
    There is a mix of both. For movement the agents find the best strategy, but the strategy for task assembly remained similar to the $14^{th}$ edition forming groups of agents around a goal cluster, but this time with more groups of agents. 
    
    \item \q{Did you implement any strategy that tries to interfere with your opponents?}\\
    Yes, we had a dedicated role (called bully) that would pursue goal clusters that were unoccupied by our team and clear any blocks from the opponents.
    
    \item \q{How do your agents decide which tasks to complete?}\\
    Our agents decide which task to complete based on the available blocks that are positioned around an active goal cluster.
    
    \item \q{How do your agents coordinate assembling and delivering a structure for a task?}\\
    Assembling is coordinated by an agent with the role of task origin which orchestrates and calls the retrievers (agents with blocks) to help it assemble a block structure. Deliverer agents are assigned to a goal cluster and remain next to a taskboard nearby, which upon receiving a signal from the task origin that a task has been selected it will accept the task at the taskboard and move to the origin position to switch places with the origin agent and deliver the task.
    
    \item \q{Which aspect(s) of the scenario did you find particularly challenging?}\\
    Because we were using an external tool for AI task planning, the most challenging aspect was to maintain good computation performance in rounds 2 and 3 (with 30 and 50 agents respectively).
  \end{description}

  \subsection{And the moral of it is \ldots}
  \begin{description}
    \item \q{What did you learn from participating in the contest?}\\
    Caching the plans was a really interesting solution, but to be effective we should have executed it a lot more times with 30 and 50 agents.
    
    \item \q{What advice would you give to yourself before the contest/another team wanting to participate in the next?}\\
    Debugging and testing the code as much as possible is better than adding new features.    
    
    \item \q{What are the strong and weak points of your team?}\\
    The automated task planning component is both the strong and weak point of our team. When we can call the planner and remain inside the deadline for sending an action it works perfectly (e.g., round with 15 agents), but when the necessary plans are not cached some agents do not send their actions on time and that crashes our team (e.g., rounds with 30 and 50 agents).
    
    \item \q{Where did you benefit from your chosen programming language, methodology, tools, and algorithms?}\\
    Coordination was very simple to achieve with Moise, and agent programming in Jason is straightforward if there is some previous knowledge of Belief-Desire-Intention systems. The planner was essential for reducing the reasoning load of the agents.
    
    \item \q{Which problems did you encounter because of your chosen technologies?}\\
    Debugging our system was hard due to the use of multiple languages and tools.
    
    \item \q{Did you encounter previously unseen problems/bugs during the contest?}\\
    Running it locally, even for 30 and 50 agents would work most of the time. During the contest it mostly did not work, and we believe this could have been caused by the extra latency with the communication to the server.
    
    \item \q{Did playing against other agent teams bring about new insights on your own agents?}\\
    Yes, mostly to show us where our bugs were happening. For the most part we tested only with a single team, since testing with two would be very strenuous to the computer running them, and running it remotely was not feasible for long testing sessions.
    
    \item \q{What would you improve (wrt. your agents) if you wanted to participate in the same contest a week from now (or next year)?}\\
    If the number of agents is this high again we would have to extensively test the cache strategy or stop trying to use external tools, since the deadline is too short to make proper use of them.
    
    \item \q{Which aspect of your team cost you the most time?}\\
    Adapting our past strategies to work with a large number of agents.    
    
    \item \q{What can be improved regarding the contest/scenario for next year?}\\
    More actions that support interactions between opposing teams would be interesting (either in the next version of this scenario, or in a new scenario).
    
    \item \q{Why did your team perform as it did? Why did the other teams perform better/worse than you did?}\\
    Our team improved in performance on the second day, despite no changes to the code. This happened because we were able to populate the cache of plans with real data from the previous matches. 
    
    \item \q{If you participated in the ``free-for-all'' event after the contest, did you learn anything new about your agents from that?}\\
    The more agents there are in the map the more problems and bugs we find.
    
  \end{description}
\section{Conclusion}
\label{sec:conclusion}

In this paper we have described the new strategies that we developed to handle the extensions of the ``Agents Assemble'' scenario in the $15^{th}$ MAPC. These strategies all contributed to our performance and resulted in our team (MLFC) obtaining 2$^{nd}$ place in the contest. The increase in the number of agents made it particularly hard to adapt our strategies, namely the use of the automated planners, since instead of up to 10 agents calling an instance of the planner (potentially in the same step), now we had to cope with up to 50 agents. The use of a plan cache was a very effective solution for this problem; however, we underestimated how much the cache could differ depending on the parameters of the round (15, 30, and 50 agents). The use of the new `bully' role contributed a lot to keeping the score of opposing teams in check by delaying and sometimes even annulling their attempts at assembling blocks and delivering tasks.

Most of our tests were with 15 agents, which did not build a sufficient pool of cached plans for rounds with 30 and 50 agents. This was demonstrated by our poor performance in the first day, when many of our agents could not access a cached plan and had to call a new instance of the planner, and thus, overload the processor and subsequently missing its deadline to send an action to the server. A known issue of our base code is that if our agents miss their deadline (i.e., server registers no action), then it is very likely that they will stop responding and will require to be restarted. This in turn is cascaded into another known issue, which is that in order to restart one agent we have to restart the whole team. Therefore, future extensions of our team for this scenario should consider more extensive testing in rounds with 30 and 50 agents to build a better plan cache, add fault tolerance so that agents can recover if they miss their deadline for sending an action, and add a feature that allows individual agents to be reconnected in case they are not able to recover.

\bibliographystyle{plain}
\bibliography{lfc}  

\end{document}